\documentclass[%
 reprint,
superscriptaddress,
 amsmath,amssymb,
 aps,prl
]{revtex4-2}

\usepackage{graphicx}% Include figure files
\usepackage{dcolumn}% Align table columns on decimal point
\usepackage{bm}% bold math
\usepackage[dvipsnames]{xcolor}
\usepackage{hyperref}
\usepackage[normalem]{ulem}
\usepackage{bbding}
\usepackage{cleveref}

\begin{document}
\title{Localized states with non-trivial symmetries: localized labyrinthine patterns}

\author{M.G. Clerc}
\affiliation{Departamento de F{\'i}sica and Millennium Institute for Research in Optics, 
FCFM, Universidad de Chile, Casilla 487-3, Santiago, Chile}
\author{ S. Echeverr\'ia-Alar}%
\affiliation{Departamento de F{\'i}sica and Millennium Institute for Research in Optics, 
	FCFM, Universidad de Chile, Casilla 487-3, Santiago, Chile}
\author{M. Tlidi}
\affiliation{Facult\'{e} des Sciences, Universit\'{e} libre de Bruxelles (U.L.B), CP. 231, 1050 Brussels, Belgium}

%


%\%affiliation{
%\affiliation{
 %Third institution, the second for Charlie Author
%\affiliation{%


\date{\today}% It is always \today, today,

\begin{abstract}
The formation of self-organized patterns and localized states are ubiquitous in Nature. Localized states containing trivial symmetries such as stripes, hexagons, or squares have been profusely studied.
 Disordered patterns with non-trivial symmetries such as labyrinthine patterns are observed in different physical contexts. Here we
report stable localized disordered patterns in spatially extended dissipative systems. 
These 2D and 3D localized structures consist of an isolated labyrinth embedded in a homogeneous steady state.  
Their partial bifurcation diagram allows us to explain this phenomenon as a manifestation of a pinning-depinning transition.  
We illustrate our findings on Swift-Hohenberg type of equations and other well-established 
models for plant ecology, nonlinear optics, and reaction-diffusion systems.
\end{abstract}

\maketitle

Spatiotemporal patterning resulting from a symmetry-breaking instability is a central issue in almost all 
driven far from equilibrium systems  \cite{Glansdorff_Prigogine,Cross-Hohenberg, Murray}. Localized structures, dissipative solitons, and localized patterns belong to this field of research. 
They consist of one or more regions in one state surrounded by a  region in a qualitatively different 
state \cite{Purwins2010, TlidiPTRS-14,Residori-Clerc, Knobloch2015,Lugiato2015}.
Spatial localization appears not only in nonlinear systems but can occur 
in linear ones such as Anderson localization
that arises in inhomogeneous systems \cite{Anderson}. 
Localized states appear in other classes of experimentally 
relevant systems such as nonlinear optics and photonics. Spatial localized patterns possess potential applications to  all-optical control of light, optical storage, and
information processing  \cite{TlidiPTRS-14,Lugiato2015}. 

\begin{figure}[b]
\centerline{
\includegraphics[width=8.5 cm]{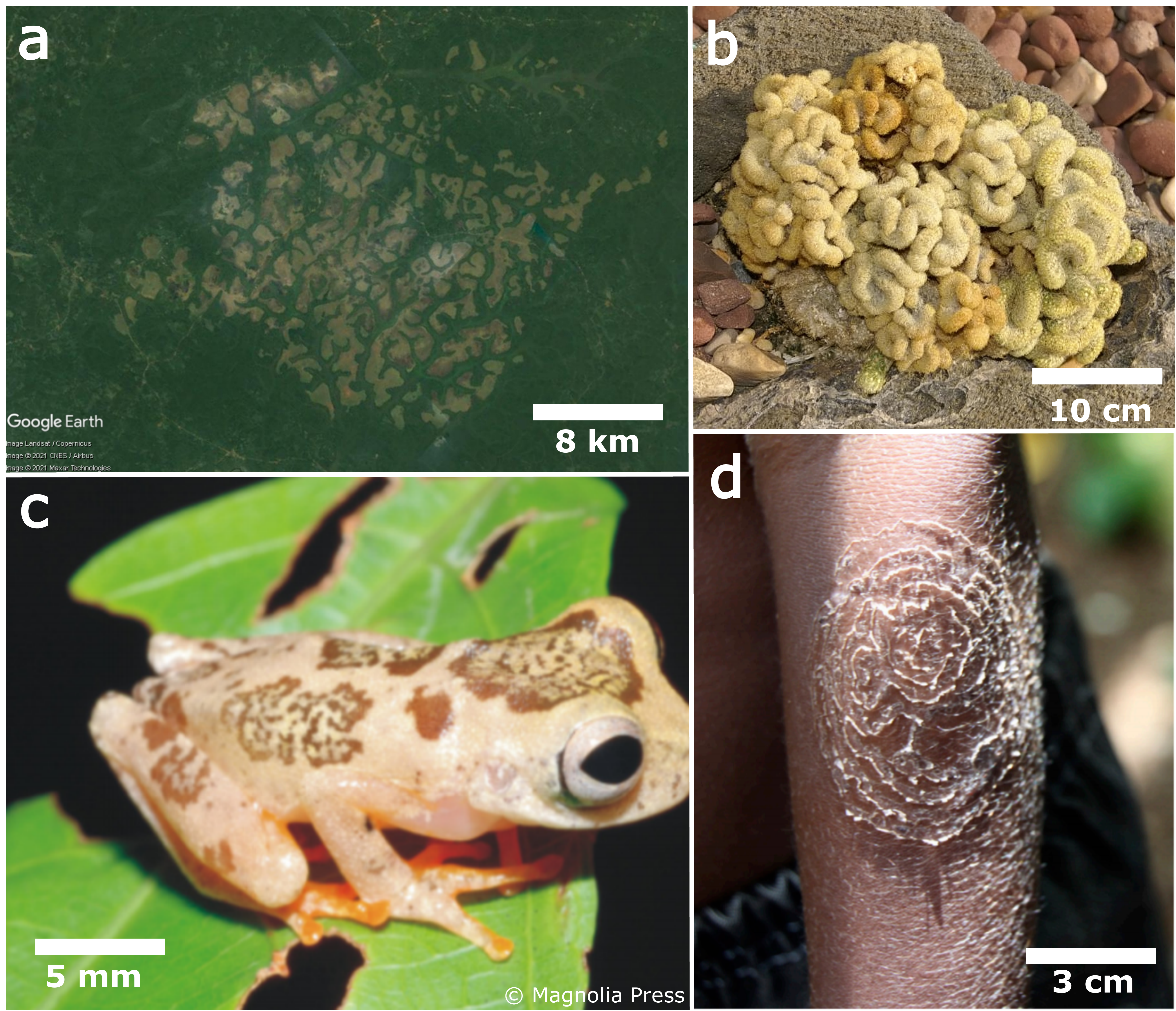}
}
  \caption{ Snapshots of localized labyrinthine patterns in natural systems. 
 	 (a) Irregular distribution of vegetation embedded in a uniform vegetated cover observed in central Cameroon using Google Earth software (with ground 
 		coordinate 3$^\circ$58'22.70" N 12$^\circ$19'05.84" E).
 		(b) Mamilaria Elongata Cristata with a contorted tissue (courtesy of David Stang). It is a localized structure in the bare soil background.
 		(c) Smudgeds composed of stripes and spots in the skin dorsum of Dendropsophus ozzyi species \cite{Orrico2014} (reproduced with permission from the copyright holder).
 		(d) Skin lessions of Tinea imbricata disease (courtesy of Michael Marks).
 }
 \label{fig:schem}
\end{figure}

Localized patterns involving trivial symmetries such as stripes, hexagons, or squares have 
been abundantly discussed and is by now fairly well understood, including their respective snaking bifurcation diagrams \cite{Coullet2002,Knobloch2015}. 
Indeed, these localized patterns involve few Fourier modes. However, localized patterns with non-trivial symmetries have 
neither experimentally observed nor documented, nor theoretically predicted. 
An example of this type of patterning phenomenon is referred to as localized labyrinthine patterns (LLP). 
They are observed in population biology, such as in vegetation populations, on  the
skin of animals or human bodies [cf.~Fig.~\ref{fig:schem}].
All these examples show an area, which is not necessarily circular, containing complex spatial structures, 
a labyrinth, and surrounded by a uniform state. In the vegetation populations, this intrigued phenomenon seems to be stationary [see the Supplementary Material \cite{Supplemental}. Extended labyrinthine patterns refer to 2D or more dimensional dissipative structures  characterized by a circular or spherical powderlike 
spectrum globally \cite{LEBERRE2002}, and
they exhibit a short-range order with a single Fourier mode \cite{CERC2020}. A power spectrum 
with a powdered ring (sphere) structure is the main characteristic of patterns with non-trivial symmetries. 
%The localized labyrinths are stationary and robust patterns since the localized area 
%never expands despite competition between plants and never shrinks despite nonlinearity and %dissipation.

In this Letter, we account for the formation of localized patterns with non-trivial symmetries in 
well established models from ecology, optics, and reaction-diffusion systems.   
We illustrate and investigate this phenomenon using a Swift-Hohenberg equation (SHE)  \cite{SH},  
which constitutes a well-known paradigm in the study of spatial periodic or
localized patterns in spatially extended systems \cite{Cross-Hohenberg}. We show that this model supports static and stable LLP. 
Considering adequate initial conditions, LLP are generated in the coexistence region between extended labyrinth and homogenous state. 
We draw the partial bifurcation diagram showing the stability domain of LLP and their pinning-depinning 
transitions, where the localized labyrinth exists as a robust solution. Free energy allows us to study the 
relative stability analysis. We show numerical evidence of 
stable three-dimensional localized labyrinthine patterns. Further within the pinning range of parameters, three LLP with different sizes are generated for a fixed value of the system parameters.

The SHE reads (\cite{SH})
\begin{equation}
\partial _{t}u=\epsilon u-u^{3}-\nu \nabla ^{2}u-\nabla ^{4}u,
\label{SHE}
\end{equation}
where the real order parameter $u=u(x,y,z,t)$  is an excess scalar field variable measuring the deviation from criticality, 
$\epsilon$ is the control parameter, and 
$\nu$ the  (anti) diffusion coefficient for (positive) negative value.  
The cubic term accounts for the nonlinear response of the system under study.
The Laplace operator $\nabla ^{2}=\partial _{xx} +\partial _{yy}+\partial _{zz}$
acts in the $(x,y,z)$-Euclidean space, and $t$ is time. 
The last term on the right-hand side, the bi-Laplacian, stands for hyper-diffusion.
Note that Eq.~(\ref{SHE}) can also be used to describe two-dimensional 
systems, where the Laplacian, bi-Laplacian and the order parameter $u$ are defined in the $(x,y)$-Euclidean space. The model equation (\ref{SHE}) can be rewritten in a variational form as $\partial _{t}u=-\delta {\cal{F}}/\delta u$, where $ {\cal{F}}$ is a Lyapunov functional or a free energy 
\begin{equation}
{\cal{F}} =\int \frac{dxdydz}{2} \left( -\epsilon u^{2} +\frac{u^{4}}{2} -\nu(\nabla u)^{2}+ (\nabla^{2} u)^{2} \right).
\label{Lyan}
\end{equation}
The variational structure of the SHE (\ref{SHE}) indicates that only stationary solutions such as uniform states, 
spatially periodic, or localized patterns are possible.  
The SHE  (\ref{SHE}) is a well-known paradigm for the study of periodic and localized patterns, 
and has been first derived  in hydrodynamics \cite{SH}, and later in other fields of natural science, 
such as chemistry \cite{Hilali}, and nonlinear optics \cite{Mandel}. 
In the last decade, it has been established that Eq.~(\ref{SHE}) had already been
constructed by Alan Turing, but was unpublished \cite{Turing}.

Other real SHE has been derived for out of equilibrium systems  \cite{NVSHE,Kozyreff2007}  
\begin{equation}
\partial _{t}u=\eta-\epsilon u-u^{3}-(\nu-b u) \nabla ^{2}u-\nabla ^{4}u-c (\nabla u)^{2},
\label{SHENV}
\end{equation}
where $\epsilon$ and $\eta$ are control parameters, $\nu$, $b$ are diffusion parameters, and $c$ is the nonlinear advection strength.
The presence of nonlinear diffusion and nonlinear advection 
terms, $u\nabla ^{2}u$ and $(\nabla u)^{2}$, render Eq.~(\ref{SHENV}) non-variational.
In general, this equation does not admit a Lyapunov functional. 
\begin{center}
 \begin{figure}
\centerline{
\includegraphics[width=8.5 cm]{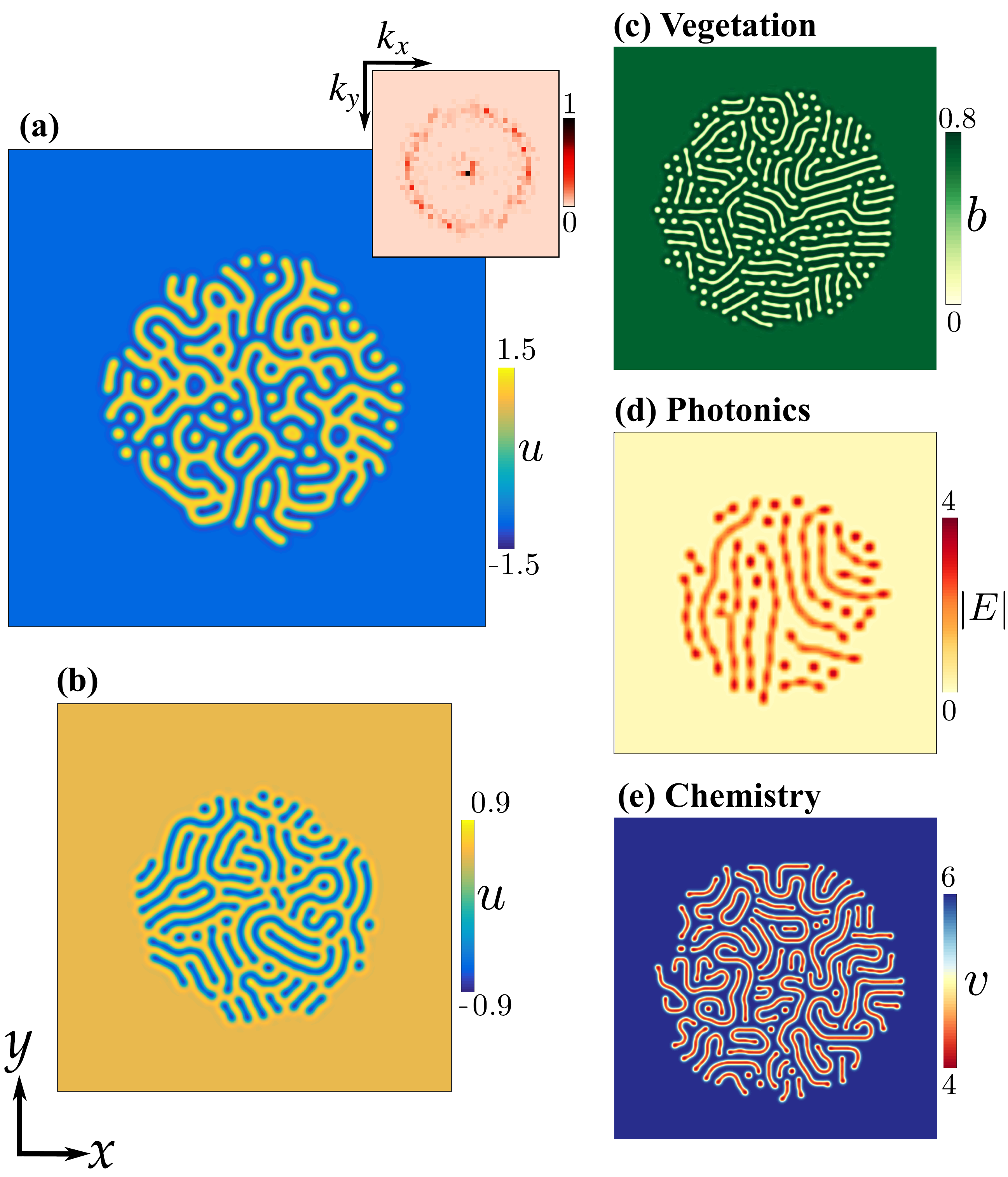}
}
 \caption{Stationary localized labyrinthine patterns obtained in different pattern forming models: (a) SHE (\ref{SHE}) ($\epsilon=1.17$, $\nu=1$). (b) generalized SHE (\ref{SHENV}) ($\epsilon=0.2$, $\nu=1$, $\eta=-0.06$, $b=0.1$, $c=0.1$). (c) Non-local vegetation. (d) Passive diffractive resonator. (e) reaction-diffusion. See the Supplementary Material for more details of models and parameters used in (c), (d), and (e) \cite{Supplemental}. The right-upper inset in (a) shows the powderlike ring specturm of the LLP in the SHE model (\ref{SHE}). All the localized structures fulfill the definition of labyrinthine patterns (see Supplementary Material \cite{Supplemental} for details). }
 \label{fig2DLP}
\end{figure}
\end{center}

Whether a SHE model is variational or not, 
numerical simulations of both models, Eqs.~(\ref{SHE},\ref{SHENV}), with periodic boundary conditions show evidence of stable stationary localized 
labyrinthine patterns [see Figs.~\ref{fig2DLP}(a) and \ref{fig2DLP}(b)]. 

To obtain localized labyrinthine patterns, the initial conditions consist of a circular area of a stable labyrinthine pattern in the center of the simulation box, embedded in a uniform background.  
The evolution towards equilibrium starts with a quick adjustment of the interface mediated 
by the curvature of the stripe patterns; then, there is an accommodation of the stripe patterns in the bulk.
To the end, some retraction of stripes in the interface takes place 
(cf.~Video 1 and the stabilization of LLP in the Supplementary Material \cite{Supplemental}). The localized region is not perfectly circular, 
containing finite segments of deformed stripes separated by spots of the same width. 
These finite-size stripes can be interconnected or not. They 
support all stripe orientations along the motionless interface separating the labyrinth 
to the homogeneous steady state, as shown in Fig.~\ref{fig2DLP}.  
Besides, the formation of 
LLP in the above scalar model equations in the form of SHE, additional models are 
also considered that are experimentally relevant: (i) a generic interaction redistribution 
model describing vegetation pattern formation which is an integrodifferential model equation. 
This simple modeling approach based on the interplay between short-range and long-range interactions
governing plant communities captures localized labyrinthine pattern as shown in Fig.~\ref{fig2DLP}(c). 
(ii) broad area photonics devices such as nonlinear resonators subjected to a coherent injected beam 
[see Fig.~\ref{fig2DLP}(d)]. In this case, the resulting equation is a complex Ginzburg-Landau type equation. 
Finally, a reaction-diffusion model for chemical dynamics, also supports LLP as shown in Fig.~\ref{fig2DLP}(e).  
The description of these models and the values of the parameters are provided 
in the Supplementary Materials \cite{Supplemental}. Similar solutions when using Dirichlet and Neumann 
boundary conditions are observed (see the Supplementary Material \cite{Supplemental} for details).
\begin{center}
	\begin{figure}[b]
		\centerline{
			\includegraphics[width=6. cm]{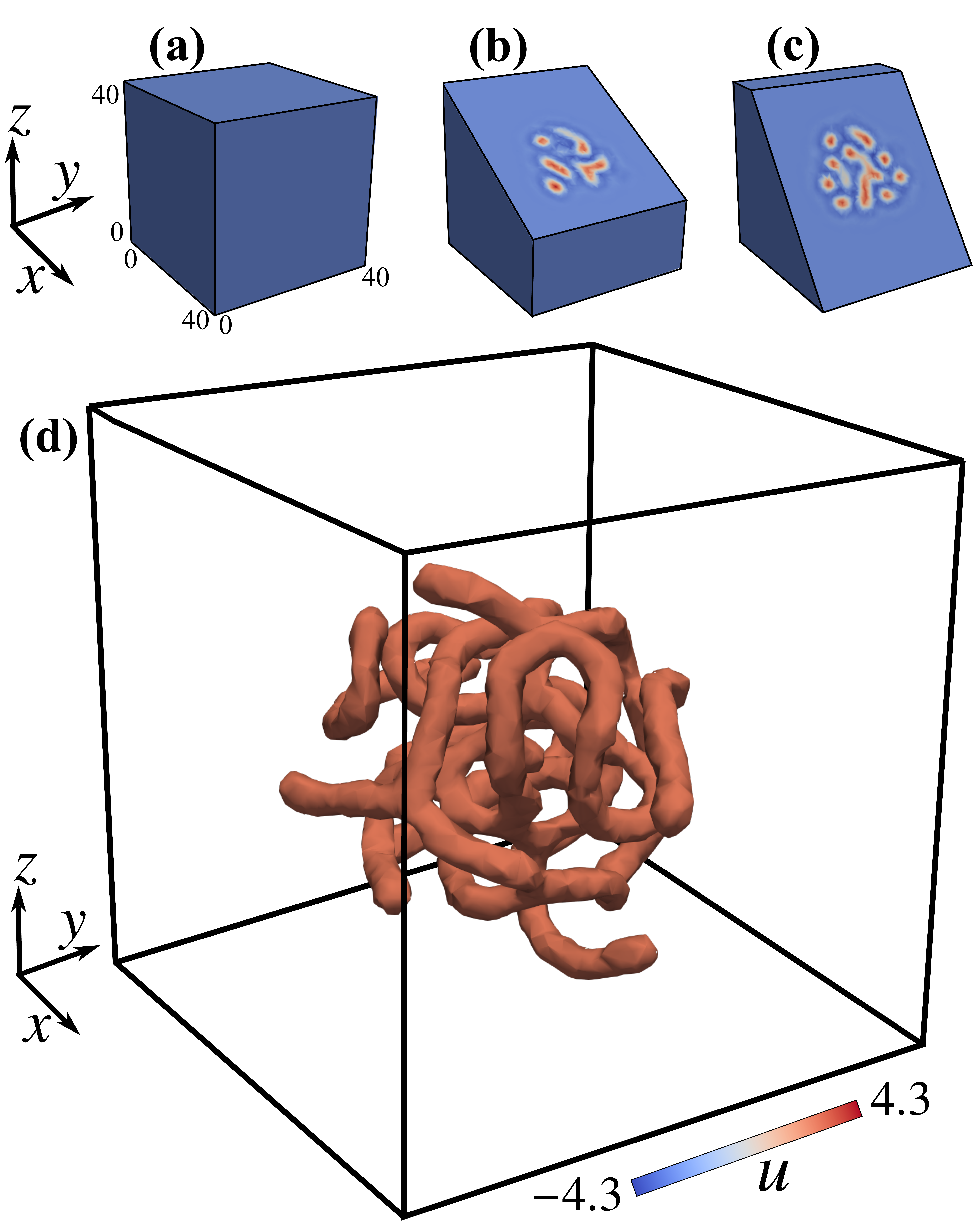}
		}
		\caption{Three dimensional localized labyrinthine pattern solution 
		of Eq.~(\ref{SHENV}). (a) Colormap of the full simulation, (b) and (c) are colormap slices of the localized labyrinthine pattern. 
			(d) Isosurface of the localized labyrinthine pattern with $u=1.3$. Parameters are 
			$\epsilon=4.2$, $\nu=5$, $\eta=-6.8$, $b=0$, and $c=0$. The mesh integration is $40\times40\times40$.}
		\label{fig3D}
	\end{figure}
\end{center}

The 2D LLP are robust structures in 2D systems in the various natural system, as shown in Fig.~\ref{fig:schem} and in the Supplementary Material \cite{Supplemental}. It has been shown that the Swift-Hohenberg equation supports 3D extended patterns with trivial symmetries such 
as lamellae, body-centered cubic crystals, hexagonally packet cylinders \cite{Thiele2013,Lee2017,Staliunas1998,TlidiMandel1998}, and 
localized patterns
\cite{Staliunas1998,TlidiMandel1998,Bordeu2015}. Recently, clusters of three-dimensional 
bullets forming a localized crystal with trivial symmetry have been reported  \cite{Gopalacresna2021}.
We extend this analysis to 3D non-trivial symmetry patterns and we show the existence of stable 3D localized labyrinthine patterns. 
They consist of finite size curved and connected tubes embedded in a homogeneous background. 
The width of the tubes is half of the critical wavelength at the symmetry-breaking instability. 
They are obtained by numerical simulations of the generalized SHE Eq. (\ref{SHENV}) with Neumann 
boundary conditions along $x$, $y$, and $z$ directions.  
Figure~\ref{fig3D} shows a typical 3D localized labyrinthine pattern.

\begin{center}
	\begin{figure}
		\centerline{
			\includegraphics[width=8.5 cm]{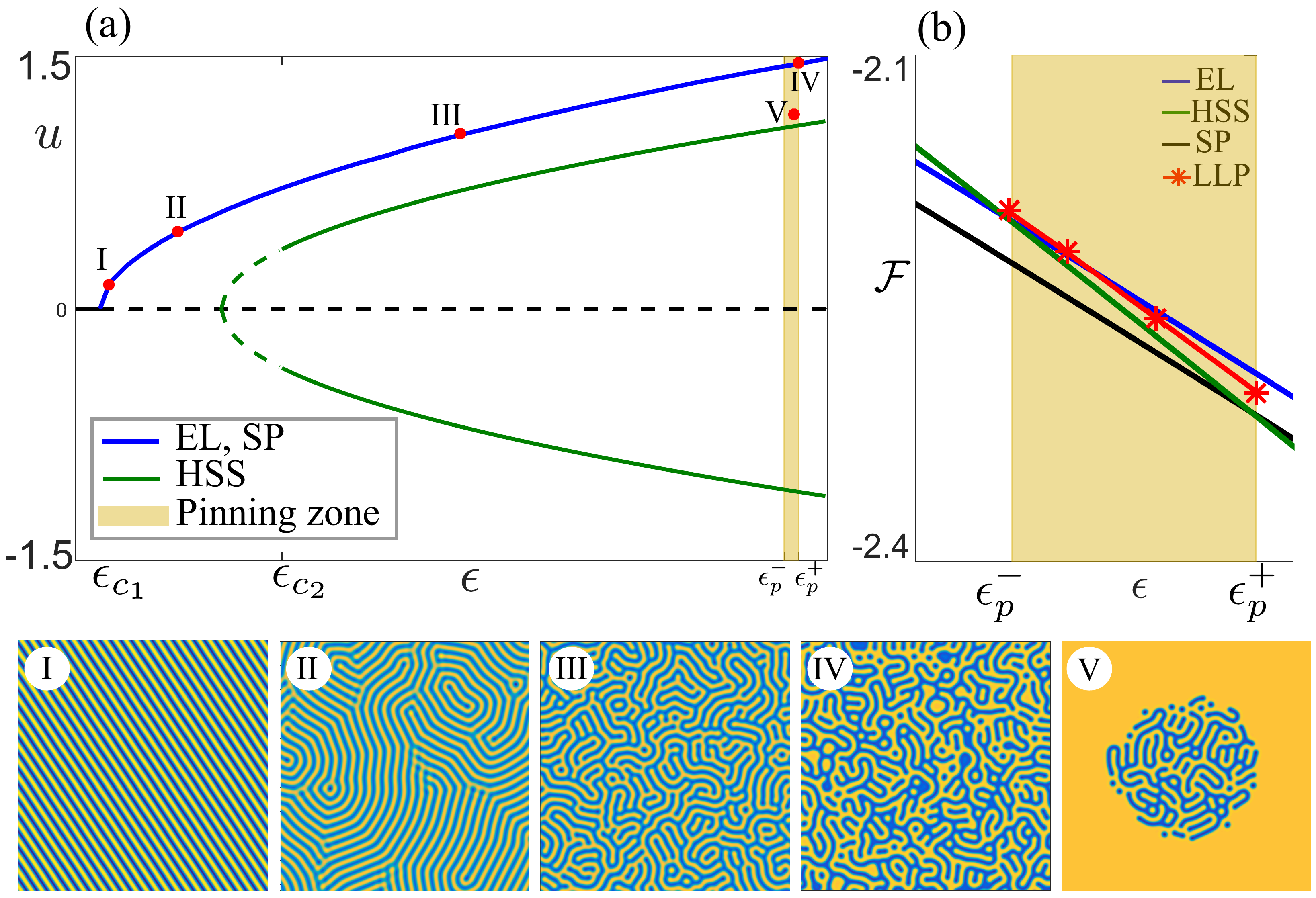}
		}
		\caption{
			Bifurcation diagram of homogenous solutions (a) and relative stability analysis of LLP (b) in SHE (\ref{SHE}), with $\nu=1$. The uniform $u(x,y)=0$ state 
			suffers a Turing instability at $\epsilon_{c1}=-0.25$. The uppermost curve (blue) shows 
			the maximum $u(x,y)$ of the different equilibrium patterns [stripe (I), fingerprint-type labyrinth (II), glassy labyrinth (III), and scurfy labyrinth (IV)]. At $\epsilon_{c2}=0.125$, the non-zero homogenous states (HSS-green curves) 
			are unstable to in-homogenous perturbations. In the narrow region limited by 
			$\epsilon_{p}^{-}=1.16$ and $\epsilon_{p}^{+}=1.19$, where the extended labyrinthine pattern 
			and the uniform solutions coexist, the existence of localized labyrinthine patterns (V) is possible. Free energy ${\cal{F}}$ given by Eq. (\ref{Lyan}) is computed for the extended labyrinths (EL), 
			the stripe pattern (SP), the homogenous states 
			(HSS), and the localized labyrinthine patterns (LLP) near the pinning region.}
		\label{fig3BIF}
	\end{figure}
\end{center}

The homogeneous steady states $u_s=0$ and $u_{s\pm}= \pm\epsilon^{1/2}$ 
solutions of Eq.~(\ref{SHE}) undergo symmetry-breaking instabilities 
at $\epsilon_{c1}=-\nu^{2}/4$ and $\epsilon_{c2}=\nu^{2}/8$. At both critical bifurcations points the critical wavelength
is $\lambda_c=2 \pi/k_c= 2\sqrt{2} \pi/\sqrt{\nu}$.
Indeed, when the linear coefficient of the Laplacian is negative; $\nu>0$, the spontaneous pattern formation process becomes possible 
thanks to the appearance of a finite band of linearly 
unstable Fourier modes that triggers the appearance of spatially periodic patterns. The upper cutoff is affected by the bi-Laplacian 
term, which is always stabilizing for short distances, since  dispersion is an efficient mixing mechanism. Numerical simulations 
of the bidimensional Eq.~(\ref{SHE}) in the neighborhood of the 
critical point  $\epsilon=\epsilon_{c1}$ indicate the emergence of extended patterns, as shown in Fig.~\ref{fig3BIF}(a).
When increasing the control parameter, the sequence of symmetry-breaking transitions fingerprint, glassy, 
and scurfy labyrinthine patterns are observed \cite{CERC2020}.

Spatial confinement leading to the formation of localized patterns with nontrivial
symmetry occurs in parameter space 
($\epsilon>\epsilon_{c2}$), where extended labyrinthine patterns coexist with a homogeneous steady state. 
Within this hysteresis loop, there generally exists a so-called pinning range of parameters
\cite{Pomeau1986},  delimited by $\epsilon^{\pm}_{p}$, in which 
robust LLP can be observed [cf. Fig.~\ref{fig3BIF}(a)]. Taking advantage of the variational structure of Eq. (\ref{SHE}), 
we address the problem of the relative stability analysis.  We evaluate numerically $\mathcal{F}$, associated with uniform states, extended labyrinth, and LLP. Figure \ref{fig3BIF}(b) summarizes the results. 
These equilibria correspond to a local or global minimum of Lyapunov functional $\mathcal{F}$ given by Eq.~(\ref{Lyan}).
From Fig.~\ref{fig3BIF}(b), we see that the LLP are more robust than the extended labyrinth but less stable than the homogeneous steady state. 
The localized labyrinth solutions are stable only in the pinning range of parameters  $\epsilon^{-}_{p}<\epsilon<\epsilon^{+}_{p}$. 
The localized labyrinths are stationary and robust patterns since their localized area 
never expands despite diffusion and never shrinks despite nonlinearity and dissipation.

\begin{center}
	\begin{figure}
		\centerline{
			\includegraphics[width=7. cm]{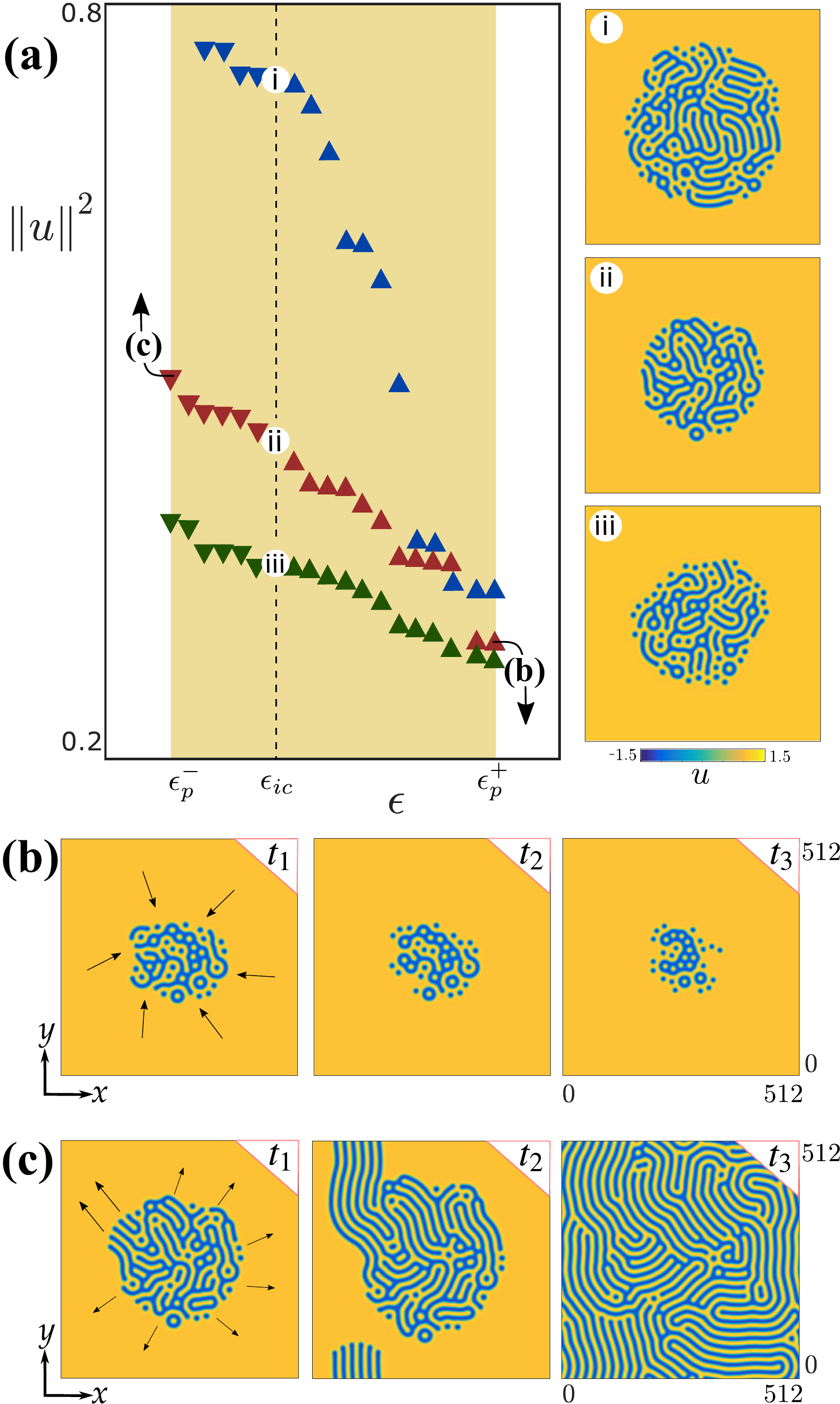}
		}
	\caption{
		Three stable branches of LLP and pinning-depinning transitions in the SHE model (\ref{SHE}). (a) Plot of $||u||^{2}$ for three 
			initial conditions (i, ii, and iii) with different
			sizes (see Fig. 3 in the Supplementary Material \cite{Supplemental} for details).
			The upward (downward) triangles account for the increasing (decreasing) of $\epsilon$, starting from $\epsilon_{ic}= 1.17$. (b) Temporal sequence of the pinning-depinning
			transition when crossing $\epsilon_{p}^{-}$=1.16, and (c) $\epsilon_{p}^{+}=1.19$.}
	
		\label{fig4BIF}
	\end{figure}
\end{center}

Localized patterns with trivial symmetry (stripes and hexagons) are organized into a complex diagram following a homoclinic snaking bifurcation \cite{Knobloch2015}. This type of diagram is obtained by a continuation method. The symmetries of the localized structures are relevant for the convergence of this algorithm. However, in the case of LLP there is a lack of continuation algorithms to characterize the full bifurcation diagram. To figure out the existence region of stable LLP, we performed direct numerical simulations of Eq.~(\ref{SHE}). Figure \ref{fig4BIF}(a) summarizes the results, where we plot 
	\begin{equation}
		||u||^{2}=\dfrac{1}{L_{x}L_{y}}\int_{0}^{L_{x}}\int_{0}^{L_{y}}(u(x,y)-u_{s+})^{2}dxdy,
	\end{equation}
as a function of the bifurcation parameter. The full bifurcation diagram can be complex, so we display only 
three branches of LLP obtained with different initial conditions shown in the insets (i), (ii), and (iii) of Fig.~\ref{fig4BIF}(a). 
The maximum amplitude of the three LLP 
is the same, but they have different size. Varying $\epsilon$ from these initial conditions, we obtain the 
three branches shown in Fig.~\ref{fig4BIF}(a). Whatever the initial condition, when increasing the bifurcation parameter 
the LLP decrease in size, mediated by shrinking of fingers and accommodation
of defects [see the Supplementary Material \cite{Supplemental} for details]. All LLP disappear
close to $\epsilon>\epsilon_{p}^{+}$ and the system exhibit a transition towards a mixture of circular localized peaks and dips. Figure~\ref{fig4BIF}(b) illustrates this transition, during which we observe the contraction of fingers, which transform to circular peaks or dips. This process correspond to the inverse of the invagination of localized structures \cite{Bordeu2015l}. Starting from the initial conditions shown in the insets (i), (ii), and (iii) of Fig.~\ref{fig4BIF}(a) and decreasing the bifurcation parameter, we observe an increase in the size of the LLP. Further decreasing $\epsilon<\epsilon_{p}^{-}$, we observe a transition to an extended fingerprint-like labyrinthine pattern. This depinning transition mediated by front propagation has the tendency to reduce the number of circular spots and dips, and enhance the invagination process as illustrated in Fig.~\ref{fig4BIF}(c).

By varying the control parameter within the pinning region delimited by $\epsilon_{p}^{-}$ and $\epsilon_{p}^{+}$, 
we see that for a fixed $\epsilon$, and near $\epsilon=\epsilon_{p}^{-}$ the sizes of the coexisting LLP are different. However, close to $\epsilon=\epsilon_{p}^{+}$ 
the system reaches more or less the same size. 
We stress that the position of LLP and their size depends on the initial conditions, 
and the maximum of the coexisting LLP  is essentially
constant for fixed values of the system parameters. The number of coexisting LLP with different sizes can be 
much larger than the three branches shown in the bifurcation diagram displayed in Fig.~\ref{fig4BIF}(a).  

The localized patterns with trivial symmetries have a well established bifurcation diagram based on continuation methods. However, when dealing with localized patterns with non trivial symmetries, there are no available algorithms for the continuation to handle this problem. Whether the localized labyrinthine 
patterns present a homoclinic snaking bifurcation diagram or not remains an open question. The plausibility of spatial varying parameters 
can be responsible for complex localized patterns. However, our result opens a novel possibility 
of localized patterns with non-trivial symmetries even in homogenous and isotropic systems.

\smallskip
\begin{acknowledgments}
	MGC thanks for the financial support of  ANID--Millennium Science Initiative Program--ICN17\_012 (MIRO) and FONDECYT
	projects 1210353. S.E.-A. acknowledges the financial support of ANID by 
	Beca Doctorado Nacional 2020-21201376. A part of this work was supported by 
	the "Laboratoire Associ\'{e}  International" University of Lille - ULB on "Self-organisation 
	of light and extreme events" (LAI-ALLURE). M.T acknowledges financial support from the 
	Fonds de la Recherche Scientifique FNRS under Grant CDR no. 35333527 "Semiconductor optical comb generator".
\end{acknowledgments}

\end{document}